\pdfoutput=1
\documentclass{article}

\usepackage{style}
\usepackage[utf8]{inputenc}
\usepackage[T1]{fontenc}
\usepackage{microtype}
\usepackage{newtxtext}
\usepackage{fancyvrb}
\usepackage{enumitem}

\usepackage{hyperref}
\usepackage{url}
\usepackage{doi}
\usepackage{orcidlink}

\hypersetup{
    colorlinks=true,
    linkcolor=blue,
    filecolor=blue,
    urlcolor=blue!70!black,
    citecolor=blue,
    pdfborder={0 0 0}
}

\usepackage{booktabs}
\usepackage{amsfonts}
\usepackage{amsmath}
\usepackage{tabularx}
\usepackage{graphicx}
\graphicspath{{figures/}}

\usepackage{natbib}
\setcitestyle{square,numbers,sort&compress}

\DeclareRobustCommand{\propt}[1]{\hypertarget{prop:#1}{P#1}}
\DeclareRobustCommand{\prop}[1]{\hyperlink{prop:#1}{P#1}}

\newcommand{\yes}{$\bullet$}
\newcommand{\half}{$\odot$}
\newcommand{\no}{$\circ$}

\newcommand{\formatkeywords}{%
  \begingroup
  \def\tempdo{0}%
  \def\do##1{\if\tempdo0\else\ \and\fi##1\def\tempdo{1}}%
  \expandafter\docsvlist\expandafter{\keywordslist}%
  \endgroup
}

\newcommand{\firstpagefooter}[1]{%
    \fancypagestyle{footer}{%
        \fancyhf{}%
        \fancyfoot[C]{%
            \small\color{black!60}%
            \begin{minipage}[t]{\textwidth}%
                \vspace{-1.5cm}%
                \raggedright%
                \textcolor{black!30}{\rule{2cm}{0.3pt}}\\[0.2em]%
                #1%
            \end{minipage}%
        }%
    }%
}

\newcommand{\doctitle}{Clear Human Web: Where a Person Vouches for Every Client}
\newcommand{\docauthor}{Lorenzo Porcelli}
\newcommand{\docsubject}{cs.CR, cs.NI, cs.CY}
\newcommand{\keywordslist}{proof of personhood, Privacy Pass, EU Digital Identity Wallet}

\newcommand{\firstdraft}{22 May 2026}
\newcommand{\lastrevision}{31 October 2026}
\firstpagefooter{First draft: \firstdraft; Last revision: \lastrevision.}

\title{\doctitle}

\author{\orcidlink{0009-0004-2539-2431}\,Lorenzo Porcelli}
\affil{Independent researcher}

\hypersetup{
  pdftitle={\doctitle},
  pdfauthor={\docauthor},
  pdfsubject={\docsubject},
  pdfkeywords={\keywordslist},
  colorlinks=true,
  linkcolor=blue,
  citecolor=blue
}

\begin{document}
\maketitle
\thispagestyle{footer}

\begin{abstract}
  The web cannot tell whether a request comes from a person or from a program. Sites have checked the network address and puzzles in its place, and both have given way: automated traffic exceeds human traffic, and puzzles are solved for pay and by agents. This paper proposes the Clear Human Web, the part of the public web that sites open only to clients a person vouches for. A person's wallet presents their person identification data (PID) to one issuer, the only party that registers as a relying party, and receives a bounded number of blind-signed tokens, proofs of personhood that carry no attribute. The browser spends one token at each site that runs the gate, and the site learns that a valid token of a trusted issuer was spent, and nothing else. No field is added to the PID, no protocol is introduced, no site registers, and the publisher of a site holds the same proof as its readers. We define the Clear Human Web through five properties and seven adversaries, and state which properties the construction satisfies against each adversary. The identities a person can hold on one site in an epoch are bounded to $k$. The bound holds once the wallet limits a person to one pseudonym at the issuer. In the optimum $k=1$, and a zero-knowledge proof of PID possession and a pseudonym per site replace the tokens.
\end{abstract}
\keywords{\formatkeywords}

\section{Introduction}  \label{sec:introduction}

A website sees a network address and a client, and cannot tell whether a person stands behind the request. For most of the web's history the address was counted as one user. Where that was not enough, a puzzle was put in the way, on the assumption that a person solves it and a program does not. Neither substitute holds any longer. Automated traffic was more than 53\% of all web traffic in 2025, against 47\% for humans~\cite{imperva2026badbot}. On Cloudflare's network, as of December 2025, humans generated 47\% of HTML requests and bots other than AI crawlers 44\%~\cite{cloudflare2025radar}. The puzzle has fallen too. A program built to solve CAPTCHAs passes every reCAPTCHAv2 challenge with no more attempts than a person needs~\cite{plesner2024recaptcha}, and another exceeds human accuracy on most visual-reasoning benchmarks~\cite{viper2026}. Across seven solver services and six agents, challenge-based defences are broadly ineffective~\cite{broken_gates_2026}. Agents still fail where a realistic interface and a valid trace of actions are demanded~\cite{hll2026}, but an adaptive solver recovers within months from each new defence~\cite{salman2026captcha}. Meanwhile the puzzles have grown harder for the persons who must solve them~\cite{searles2023captcha}.

The problem has three faces, one for each party on the web. Whoever publishes has no fence. The Robots Exclusion Protocol is a request that crawlers are asked to honour, with no force of access authorisation~\cite{rfc9309}. The user agents most often disallowed by sites are those of AI crawlers, and the same crawlers return as little as one visitor for hundreds of thousands of pages fetched~\cite{cloudflare2025radar}. A publisher wants to keep out only the machines nobody declares, and to admit persons and the declared crawlers it chooses.

Whoever measures has no count. A number of readers is not a number of persons: Meta itself estimates that about 10\% of its monthly active users are duplicate accounts and about 4\% false ones~\cite{meta2024tenk}. A reader certified as a person is a metric that does not exist.

Whoever reads has no proof. By mid-2025 roughly 35\% of newly published websites were classified as AI-generated or AI-assisted~\cite{dolezal2026aiweb}. People cannot tell. In one study 4,600 participants were unable to detect self-presentations written by a language model~\cite{jakesch2023heuristics}. In another, a model prompted to adopt a humanlike persona was judged to be the human 73\% of the time in a three-party Turing test~\cite{jones2026turing}. Across 48 markets, 62\% of respondents are concerned about what is real and what is fake on the internet~\cite{reuters2026dnr}. A reader wants to know that persons stand behind what they read, and that they read among persons. The three faces have one cause, a web that cannot see the person, and call for one cure: a check that a person stands behind the client, which reveals nothing else.

Attesting a person has been tried in two ways. Scanning the iris~\cite{worldid} puts a biometric at the root. Anchoring the person to a scarce relationship such as an account or a phone number~\cite{pact2026} rests on a signal that costs an attacker what it costs a user. The European Union (EU) offers a different root. Under Regulation (EU) 2024/1183, which amends the eIDAS regulation, every Member State issues person identification data (PID) to a wallet held by its residents~\cite{eidas2}. The PID set is required to be unique for the Member State~\cite{cir2024_2977}. The Commission's own paper states that proof of possession of a PID would be a sufficient indication that the user is a human being~\cite[\S4.1.2]{arf_topic_g}. What is missing is the path from that proof to any site on the web: a PID presented to a site is linkable across sites~\cite[\S10.1]{sdjwt_ietf}, and a site that verifies it must register under the regulation~\cite{cir2025_848}.

This paper proposes the Clear Human Web, a part of the public web where every client is vouched for by a person. A site that joins it gains one fact, that a person stands behind each client. The construction has three steps. A person's wallet shows the PID to an issuer, which verifies it and hands back a bounded stock of anonymous proofs. The person's browser spends one proof at each site it visits, when the site asks for it. The site checks that the proof comes from an issuer it trusts and opens the door; it does not learn who the person is, and cannot tell that two visits come from the same person. Nothing changes in the PID, in the protocols of the web or in the obligations of the sites. The publisher of a site presents the same proof, so a person also stands behind every site. In the theoretical optimum a person proves possession of the PID without showing it, appears under one name per site, and needs no issuer; each piece exists on paper and none can be deployed yet.

The objectives follow from the three faces. A site of the Clear Human Web admits persons and the declared machines it chooses, and excludes the machines nobody vouches for. A person counts as at most $k$ visitors on any one site in an epoch, one in the optimum, and nobody can count persons across sites. The publisher of a site is a person as well, by the same proof. All three hold without the site, the issuer or the state learning who reads what. The analysis shows that the construction meets these objectives against seven adversaries, with two reserves: the network address, which only Tor removes, and the bound per person, which waits for a wallet that limits pseudonyms. The Clear Human Web bounds identities and says nothing about content: a page a person vouches for may be written by a machine. 

Section~\ref{sec:related} places the construction among the designs closest to it. Section~\ref{sec:design} translates the objectives into five properties, seven adversaries and the optimum, lists the existing pieces, and gives the construction that can be deployed with the wallets the regulation requires. Section~\ref{sec:analysis} states which properties hold against which adversary and bounds the identities a person can hold on a site. Section~\ref{sec:discussion} states what the construction does not guarantee, the three steps from it to the optimum, and what is open.

\section{Related Work}  \label{sec:related}

The Clear Human Web lets a site check that a person holding an identity document vouches for the client, and learn nothing else about them. Table~\ref{tab:comparison} compares the closest designs by what they attest, who attests it, what a site learns, and where each stops.

\begin{table}[h]
  \centering
  \scriptsize
  \renewcommand{\arraystretch}{1.2}
  \renewcommand{\tabularxcolumn}[1]{m{#1}}
  \caption{Designs closest to the Clear Human Web: what is attested, by whom, what a site learns from a visitor, and where each stops short of a web for persons.}
  \label{tab:comparison}
  \begin{tabularx}{\linewidth}{@{}>{\raggedright\arraybackslash}m{2.6cm}XXXX@{}}
    \toprule
    Design & What is attested                                                                          & By whom & What a site learns & Where it stops \\
    \midrule
    Challenge-based detection~\cite{motoyama2010captcha,broken_gates_2026}
           & A solved puzzle, or a behaviour judged human
           & The site or its intermediary
           & A score, and what the fingerprint carries
           & Solved for pay and by agents                                                                                                              \\
    \midrule
    Privacy Pass with device attestation~\cite{rfc9576,apple_pat}
           & A genuine device
           & The device vendor
           & A valid token
           & Attests the device only                                                                                                                   \\
    \midrule
    Biometric proof of personhood~\cite{worldid}
           & A unique human, by the iris
           & A private operator's devices
           & A zero-knowledge proof with a nullifier per context
           & A biometric as the root; enrolment by one operator                                                                                        \\
    \midrule
    Personhood credentials~\cite{adler2024phc,choudhuri2026pop}
           & A unique person, one credential per issuer
           & A trusted issuer, left unnamed
           & A zero-knowledge statement
           & A concept and a framework without a deployment                                                                                            \\
    \midrule
    PACT and MoLE~\cite{pact2026,mole2026protocols}
           & A scarce relationship: an account, a subscription, a phone number
           & The service that holds it
           & A nullifier per credential and epoch
           & Anchors cost an attacker what they cost a user; the nullifier does not depend on the site                                                 \\
    \midrule
    x401~\cite{x401spec}
           & Any credential the site names
           & The holder's wallet, to every site
           & The credential presented
           & Every site verifies, and in the EU registers; plain presentations are linkable                                                            \\
    \midrule
    EU age verification~\cite{av_blueprint_spec}
           & An age threshold
           & A dedicated provider, in batches of attestations
           & An age attestation
           & One attribute; the provider has seen every attestation it issued                                                                          \\
    \midrule
    Clear Human Web
           & A person holds a PID
           & One registered issuer, in blind tokens
           & A valid token of a listed issuer
           & Needs the wallets and one issuer registered as a relying party                                                                            \\
    \bottomrule
  \end{tabularx}
\end{table}

\paragraph{What sites check today.}
Sites have long checked the client rather than the person. The first signal was the network address, which identifies a device or a network. Users who share an address, Tor users above all, are the ones most often challenged by sites~\cite{privacypass_popets}. The challenge came second: by 2010 a market resold human labour to solve CAPTCHAs at as little as one dollar per thousand~\cite{motoyama2010captcha}, and programs now solve them too~\cite{plesner2024recaptcha,viper2026,broken_gates_2026}. The third check is the device: Privacy Pass tokens issued against an attestation by the device vendor let Apple devices skip the challenge~\cite{rfc9576,apple_pat}. All three check the client; none establishes that a person is there.

\paragraph{Attesting a person.}
Proof-of-personhood designs establish directly that a person is there, each from an enrolment of its own. World scans the iris with a device built by a private operator and presents a zero-knowledge proof with a nullifier; by its own count about 18 million people have been verified~\cite{worldid}. Idena verified humans with synchronous tests, and its keys were rented to puppeteers~\cite{Ohlhaver2024PoP}. Adler et al.\ abstract these designs into personhood credentials, at most one per eligible person and issuer~\cite{adler2024phc}, and Choudhuri et al.\ formalise them~\cite{choudhuri2026pop}. PACT and MoLE attest a scarce relationship instead, an account, a subscription or a phone number, through anonymous credentials native to the browser~\cite{pact2026,mole2026arch,mole2026protocols}. The enrolment is the hard part of each design, and where it is biometric, private, or for rent. The Clear Human Web runs none: the Member State already keeps one, as the PID set the regulation requires to be unique~\cite{cir2024_2977}.

\paragraph{Carrying the proof without the identity.}
Between the party that knows the person and the site that needs only the proof, something must carry a yes that neither can link to the other. Chaum's blind signature did this for payments~\cite{chaum1983blind,rfc9474}, and Privacy Pass does it on the web with a standard HTTP challenge, an issuer that signs without seeing and an origin that verifies offline~\cite{rfc9576,rfc9577,rfc9578}. A PID presented plainly is linkable across sites and recognisable by its issuer~\cite{sdjwt_ietf,frigo2026ecdsa}, and a malicious issuer can trace users through the values it signs in a batch~\cite{invisible_traces_2026}. The question is therefore where the wallet is asked. x401 asks it at every site, which demands the credential on an HTTP route~\cite{x401spec}. The age-verification solution asks it once, at a dedicated provider that issues batches of attestations, single-use as a plain mobile document (mdoc), to be verified by sites that need not register~\cite{av_blueprint_spec}. When documents meet sites directly the cost is measured. In the US states with mandates, one provider serves over 60\% of the sites that verify~\cite{papersplease2026}. It often requires face photos, government IDs and the site being accessed, which may reach fourth parties~\cite{papersplease2026}. The Clear Human Web follows the second route: one registered issuer between the wallet and the sites, which see blind tokens only.

\paragraph{Bounding identities.}
Sybil defences bound how many identities one party can hold in a system that does not know who its parties are. Without a bound a person could vouch for any number of clients, and the proof would be worthless. Douceur showed that without a logically centralised authority Sybil attacks are always possible, the alternative being a trusted agency that certifies identities~\cite{douceur2002sybil}. SybilGuard took the decentralised route through the social graph and stated the price of the centralised one. An authority issuing credentials unique to a human being, for instance against government-issued numbers, controls Sybil attacks easily. However, one entity trusted by every user worldwide is hard to establish, and requiring sensitive information scares users away~\cite{sybilguard2008}. The Clear Human Web pays that price where it is already paid, at the Member State, and answers both objections with the issuer. The sensitive information stops there, and a site receives tokens that carry none of it, in a bounded number per site and epoch.

\paragraph{Anonymous accountability.}
Anonymous accountability lets a site recognise or refuse a visitor it cannot identify. Nymble blacklisted anonymous users without learning their address~\cite{nymble2007}, and SyRA turns a legacy identity or personhood identifier into one unlinkable pseudonym per context, with at most one digital identity per user~\cite{syra2025}. Both presuppose a credential behind the pseudonym and say nothing about where a website would get it. The construction of this paper gives a site a session and no pseudonym. A pseudonym per site and epoch derived from the PID would be, in SyRA's terms, exactly such a pseudonym, and would let a site refuse a visitor as Nymble does.

\paragraph{Anonymity networks and declared machines.}
Anonymity networks hide who is speaking, and the newer work on automated clients declares that a machine is. Tor hides the client's network address~\cite{tor2004}. A token carries no address, so the two compose. Web Bot Auth lets automated clients sign their requests, identifying the machine~\cite{webbotauth_charter,webbotauth_httpsig}, and authenticated delegation extends OAuth 2.0 and OpenID Connect so that a person delegates and restricts an agent accountably~\cite{south2025delegation}. Both identify the machine, or its principal, to the service. The Clear Human Web composes with Tor and differs from the other two in what the service learns. A program a person vouches for is admitted within that person's quota, and the site learns only that a person stands behind it.

\section{Design}  \label{sec:design}

This section defines the Clear Human Web and its goals, lists the existing pieces it is built from and describes the construction that can be deployed today.

\subsection{Definition, Goals and Threat Model}  \label{sec:definition}

The Clear Human Web is best understood by contrast with a darknet. A darknet hides: since the web knows devices, not persons, it hides the device to hide the person behind it. The Clear Human Web adds the person to what the web can see, and its sole purpose is to reveal this one bit, whether a person stands behind a client or not. This bit lets a part of the public web emerge wherever sites choose to join it: visible to persons, invisible to undeclared machines, and visible to declared machines such as search crawlers only where a site admits them. The enabling mechanism is an HTTP challenge answered with a token that proves one thing, that a person holding an identity document vouches for the client, and nothing else: not who the person is, nor that they are the one reading.

The definition uses four terms. Persons are natural persons holding a PID; they act as readers or as publishers. Sites are web origins that serve resources, the pages and files identified by URLs. Issuers are services that verify a natural-person PID and hand out proofs of personhood. Agents are the software through which a person reads: a browser, possibly with a program behind it. The Clear Human Web is the set of resources that sites serve only to agents presenting a valid proof of personhood. A proof is valid if it derives from a PID presentation to an issuer the site trusts, and it carries no information about the person beyond the fact that one exists. A site belongs to the Clear Human Web when all the resources it serves to readers are gated in this way and its publisher holds a proof of the same kind.

The construction of the Clear Human Web must satisfy five properties.
\begin{description}[leftmargin=2.2em,labelwidth=1.6em,itemsep=1pt,topsep=2pt]
  \item[\propt{1}] \emph{Admission by persons.} Every access to a resource of the Clear Human Web is vouched for by a person holding a PID, and every site of the Clear Human Web is published by a person holding a PID.
  \item[\propt{2}] \emph{No identity.} No site, issuer or PID provider learns from the protocol who reads what, nor who publishes which site. The issuer learns only that some PID holder obtained proofs.
  \item[\propt{3}] \emph{Unlinkability.} Two visits by the same person, to different sites or to the same site in different sessions, cannot be linked by sites, by the issuer, or by sites and issuer together.
  \item[\propt{4}] \emph{Bounded multiplicity.} On any one site and within one epoch, the validity period of a proof, a person can appear as at most $k$ distinct visitors, with $k = 1$ in the optimum.
  \item[\propt{5}] \emph{Delegation tolerance.} A person may let a program act within their own quota. What is excluded is a program with no person vouching for it.
\end{description}

Consider Alice, who holds a PID and reads three sites, A, B and C. Her browser answers each site's challenge with a proof, and the publisher of each site holds a proof of the same kind. A crawler with no person behind it has none and is refused (\prop{1}). Site A learns that a person read a page, not that it was Alice; the issuer learns that some PID holder obtained proofs, not that she read A (\prop{2}). Neither A, B and C together, nor the issuer with them, can tell that their visitor is the same person (\prop{3}).

Each site counts her as one visitor: with $k = 1$ she cannot be two visitors on A. Summing the visitors of A, B and C counts her three times, and nobody can correct that sum into a count of persons without linking the sites, which \prop{3} forbids. The only global number compatible with the goals is how many persons enrolled at the issuers, which they can publish without naming anyone (\prop{4}). A person counts once on every site, and there is no census. Finally, Alice may let a program fetch pages within her own quota; the program is admitted because she vouches for it, and the same program with no person behind it is not (\prop{5}).

We constrain the design to reuse existing credentials, standards and software unchanged: no new field in the PID, no new protocol, no registration of sites. Within this constraint, we evaluate the properties against seven adversaries: (1) an honest-but-curious site; (2) colluding sites; (3) a site colluding with an issuer; (4) the state, as PID provider and as a possible issuer; (5) a renter, who sells the use of their own proofs; (6) a delegated program acting within a person's quota; and (7) a network observer.

The optimum, anonymous persons, satisfies \prop{1} to \prop{5} with $k = 1$ and no tokens at all. The agent holds a reusable zero-knowledge proof of PID possession. For each site and epoch it derives a pseudonym exclusive to that site, so a site sees each person once and recognises them nowhere else. Finally, the connection runs over Tor, so the network address says nothing. The genealogy is older than the wallet: Privacy Pass was designed for Tor users repeatedly challenged by sites~\cite{privacypass_popets}, and Nymble showed that misbehaving anonymous users can be blocked without learning who they are~\cite{nymble2007}. None of the three pieces is deployable yet, as the next subsection records. The age-verification solution shows that the token path is the one being rolled out, and the construction follows it.

\subsection{Building Blocks}  \label{sec:blocks}

The pieces available to the construction are EU law, published standards and shipped software. Table~\ref{tab:blocks} lists them with their status at the time of writing, what each contributes and where it stops.

Regulation (EU) 2024/1183 obliges every Member State to provide at least one European Digital Identity Wallet, free of charge for natural persons~\cite{eidas2}. The wallet holds the PID of its user. The implementing regulation defines one PID set for natural persons and one for legal persons, and requires the PID issued to a wallet user to be unique within the Member State~\cite{cir2024_2977}. We use the natural-person set, the one described by the PID rulebook~\cite{eudi_pid_rulebook}. The PID is thus a state-backed credential that exists once per person and that no platform issues.

The presentation protocol lets a verifier request a credential without requesting any of its attributes. The wallet then returns only what is mandatory, namely the signed credential and the proof that the device holds its key~\cite[\S6.4.1]{openid4vp}. Proving possession of a PID while disclosing nothing is therefore already expressible. The Commission's discussion paper on zero-knowledge proofs states the premise we build on. Proof of possession of a PID would suffice to indicate that the user is a human being, provided a rate limit prevents re-use~\cite[\S4.1.2]{arf_topic_g}.

Two limits remain. The first is linkability: a credential presented plainly can be recognised by verifiers across presentations~\cite[\S10.1]{sdjwt_ietf}, and its holder by an issuer that colludes with a verifier, for mdoc as for the selective-disclosure JWT (SD-JWT)~\cite{frigo2026ecdsa}. Issuing several copies of a credential, each presented once, mitigates the first linkage but not the second~\cite{arf_topic_b_batch,sdjwt_ietf}. The second limit is registration. Only a registered relying party may ask, since it needs an access certificate to authenticate itself towards the wallet before any presentation, as the Architecture and Reference Framework (ARF) of the wallet requires~\cite{eudi_arb}. Registration is governed by Implementing Regulation (EU) 2025/848, applicable from 24 December 2026~\cite{cir2025_848}, and the reference wallet already checks the registration certificate of the party that asks~\cite{eudi_ref_wallet_android}. A design in which every website asked the wallet directly would therefore require every site to be registered.

Privacy Pass supplies the transport~\cite{rfc9576,rfc9577,rfc9578}. An issuer signs tokens without seeing them, and any origin verifies them offline with the issuer's public key. The origin requests a token with a standard HTTP challenge naming the issuers it accepts, and the client answers with a token in its next request. Deployed by Cloudflare and Apple~\cite{privacypass_popets,apple_pat}, the architecture leaves the admission criterion open: its issuers attest to a solved challenge or to a device, never to a person.

The EU Age Verification solution shows how a wallet credential becomes such tokens for one attribute~\cite{av_blueprint_spec,av_annex_b_zkp}. A dedicated provider issues to the user's app batches of attestations, single-use when presented as plain mdoc, while a zero-knowledge presentation, the preferred method, may be reused within its validity period. Sites verify them against a trust list, and the specification states that relying parties requesting age verification need not register~\cite[\S2.6, \S3.4, \S4.2]{av_blueprint_spec}. A Commission Recommendation of April 2026 urges Member States to make the solution available by the end of 2026~\cite{ec_reco_av_2026}. This precedent settles two questions: a tokenised proof derived from a wallet credential is acceptable at EU level, and sites that only verify tokens need not register.

Three pieces of the optimum exist on paper. First, two Commission specifications describe zero-knowledge presentations of wallet credentials. TS13 admits any suitable proof system and defines one type, the Longfellow scheme of Frigo and shelat~\cite{eudi_ts13_zk,frigo2026ecdsa}. It proves possession of a validly signed, device-bound, unexpired and unrevoked mdoc without any change on the issuer side, in under a second on a phone~\cite{frigo2026ecdsa,eudi_ts13_zk}. Its published circuit discloses at least one attribute, and a zero-attribute variant is not described~\cite{frigo2026ecdsa}. TS14 builds on multi-message signatures~\cite{eudi_ts14_mms}. Both specifications leave pseudonyms out of scope~\cite{eudi_ts13_zk,eudi_ts14_mms}. The core library of the reference wallet offers Longfellow proofs as an option documented for proximity presentation, and the reference wallet application does not enable them~\cite{eudi_wallet_core_zkp,eudi_ref_wallet_android}.

Second, wallets must already generate, on request, a pseudonym specific to one relying party~\cite[Art.~14(2)]{cir2024_2979}. A pseudonym that bounds a person to a given number of identities within a scope has been discussed by the Commission against troll farms~\cite{arf_topic_e}. The current concept built on WebAuthn does not support it, the refinement round keeps it as a requirement and deems it necessary in the long run~\cite[\S4.1.2]{arf_topic_e_rr}, and constructions exist in the literature~\cite{mayrhofer2025pseudonyms}.

Third, MoLE, the protocol work behind PACT~\cite{pact2026,mole2026arch}, defines an endorsement in which a client proves in zero knowledge that it holds a valid mdoc from an accepted set of issuers. The endorsement outputs a nullifier unique per credential and per epoch~\cite{mole2026protocols}.

Two further pieces are orthogonal to the rest. Tor hides the network address, which the construction never touches~\cite{tor2004}. Web Bot Auth, an IETF working group, standardises signatures with which automated clients authenticate themselves to sites that primarily serve human users, the end user out of its scope~\cite{webbotauth_charter,webbotauth_httpsig}. One lane admits declared machines; the layer we construct admits persons.

What none of these pieces provides is the combination. The PID proves personhood only to registered parties. Privacy Pass carries anonymous tokens but has no personhood issuer. The age-verification design bridges the two for a single attribute. Nullifiers and rate-limited pseudonyms are specified on paper and implemented in no wallet release. The next subsection builds the combination.

\begin{table}[h]
  \centering
  \scriptsize
  \renewcommand{\arraystretch}{1.2}
  \renewcommand{\tabularxcolumn}[1]{m{#1}}
  \caption{Building blocks: what each gives the construction, how mature it is at the time of writing, and where it stops.}
  \label{tab:blocks}
  \begin{tabularx}{\linewidth}{@{}XXXX@{}}
    \toprule
    Piece & What it gives                                                                                                           & Maturity & Where it stops \\
    \midrule
    EU wallet and PID~\cite{eidas2,cir2024_2977,eudi_pid_rulebook}
          & A state-issued credential, one per natural person, free of charge
          & Regulation in force; wallets due by December 2026
          & Linkable when presented plainly; only registered parties may ask                                                                                    \\
    \midrule
    Presentation without attributes~\cite{openid4vp}
          & A proof of possession that discloses nothing
          & Final standard
          & The same credential is recognisable across presentations                                                                                            \\
    \midrule
    Access certificate\newline and registration~\cite{eudi_arb,cir2025_848,eudi_ref_wallet_android}
          & Every party that asks the wallet is identified
          & Regulation applicable from 24 December 2026; checked by the reference wallet
          & Would force every website to register                                                                                                               \\
    \midrule
    Digital Credentials API~\cite{w3c_dc_api,chrome141_dcapi}
          & Presentation mediated by the browser on the same device
          & Working Draft; on by default in Chrome; in the reference wallet
          & Not yet a Recommendation                                                                                                                            \\
    \midrule
    Privacy Pass~\cite{rfc9576,rfc9577,rfc9578,privacypass_popets,apple_pat}
          & Anonymous tokens with a standard HTTP challenge, verified offline
          & Standards Track RFCs; deployed by Cloudflare and Apple
          & No issuer attests to a person; the rate-limited variant expired as a draft~\cite{arc_crypto_draft,arc_protocol_draft}                               \\
    \midrule
    EU Age Verification\newline solution~\cite{av_blueprint_spec,av_annex_b_zkp,ec_reco_av_2026}
          & The template: batches of attestations, a trust list, no site registration
          & Specification published; roll-out urged by the end of 2026
          & One attribute, age                                                                                                                                  \\
    \midrule
    Zero-knowledge\newline presentations~\cite{eudi_ts13_zk,eudi_ts14_mms,frigo2026ecdsa}
          & A reusable, unlinkable proof of possession
          & Specifications published; proofs in under a second on a phone
          & At least one attribute disclosed; no pseudonyms; optional in the reference library and off in the reference application                             \\
    \midrule
    Pseudonyms~\cite{cir2024_2979,arf_topic_e,arf_topic_e_rr,mayrhofer2025pseudonyms}
          & One pseudonym per relying party; in the optimum, one per person and scope
          & Relying-party pseudonym mandatory; the rate-limited one a discussion paper
          & Required, but not supported by the current WebAuthn-based concept                                                                                   \\
    \midrule
    PACT and MoLE~\cite{pact2026,mole2026arch,mole2026protocols}
          & Anonymous credentials in the browser; one nullifier per credential and epoch
          & Announced; individual IETF drafts
          & Anchors are scarce signals rather than identity documents                                                                                           \\
    \midrule
    Web Bot Auth~\cite{webbotauth_charter,webbotauth_httpsig}
          & A lane for declared machines
          & Working group; first draft adopted
          & Says nothing about the end user                                                                                                                     \\
    \midrule
    Tor~\cite{tor2004}
          & Hides the network address
          & Deployed
          & Says nothing about who speaks                                                                                                                       \\
    \bottomrule
  \end{tabularx}
\end{table}

\subsection{Construction}  \label{sec:construction}

\begin{figure}[ht]
  \centering
  \includegraphics[width=.70\textwidth]{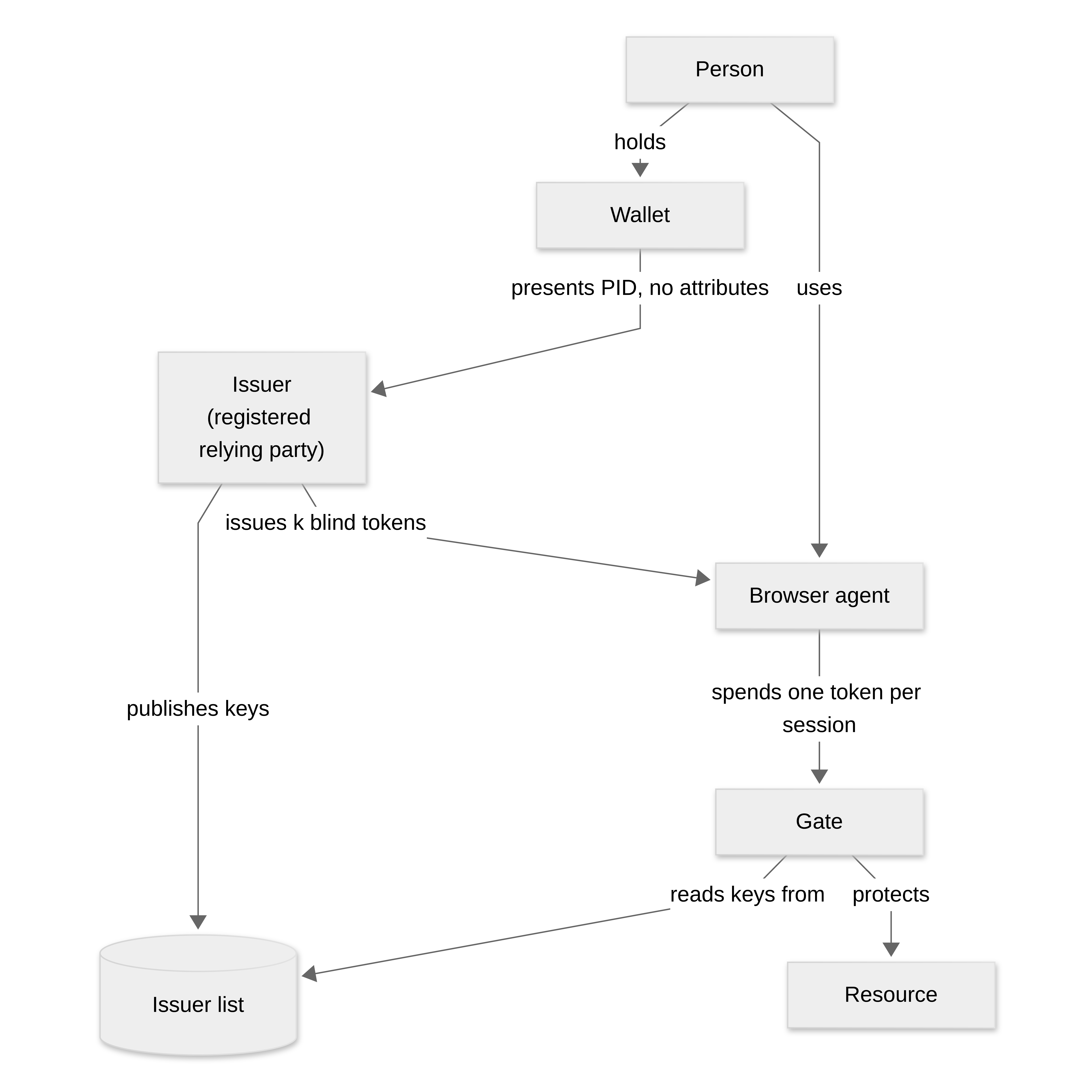}
  \caption{Components of the Clear Human Web and what each does. The issuer is the only registered relying party of the wallet. The gate verifies tokens offline with the keys it reads from the issuer list. The agent spends one token per site session.}
  \label{fig:architecture}
\end{figure}

This subsection describes a construction built from the pieces above, as they stand. The web side of the construction can be deployed now. The identity side waits for the wallets due by December 2026 and for the registration of the issuer under the regulation applicable from the same date. Nothing in it waits for a new standard. Figure~\ref{fig:architecture} shows the components and what each does. The personhood token issuer is the only party that talks to the wallet, and therefore the only registered relying party. The browser agent holds proofs for a person and answers challenges on their behalf. The gate is a reverse proxy in front of a resource. It challenges visitors and verifies proofs on the resource's behalf. The issuer list is a public document that names the issuers a site accepts and carries their public keys for the current epoch.

\begin{figure}[ht]
  \centering
  \includegraphics[width=.75\textwidth]{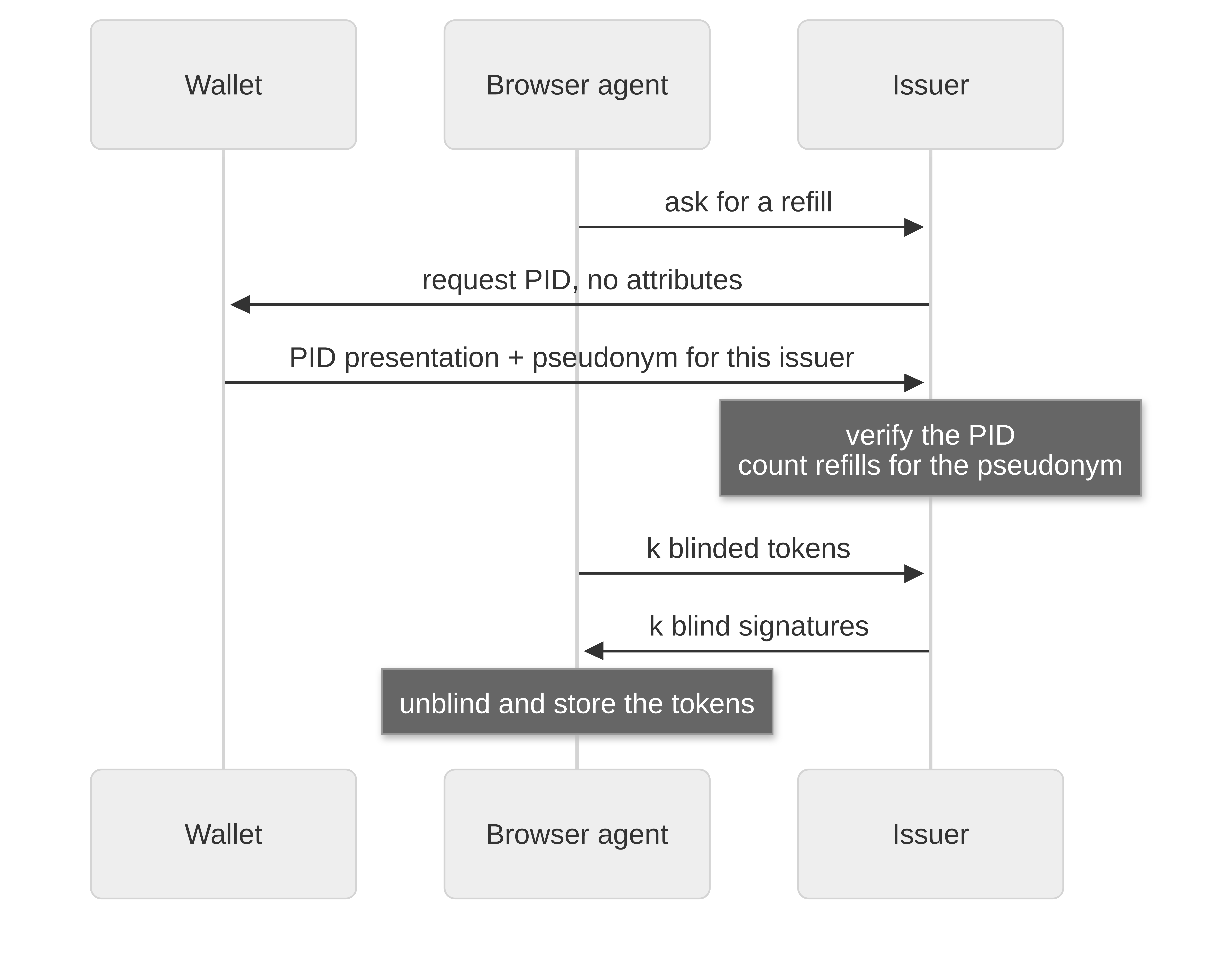}
  \caption{Refill: once per epoch, with one wallet presentation and no attribute disclosed.}
  \label{fig:refill}
\end{figure}

A person enrols once per epoch, as shown in Figure~\ref{fig:refill}. The agent asks the issuer for a refill. The issuer asks the wallet for the natural-person PID with no attributes, together with the pseudonym that the wallet generates for this issuer, through the browser on the same device or across devices~\cite{openid4vp,chrome141_dcapi,cir2024_2979}. The wallet returns the signed PID and the proof that the device holds its key. The issuer verifies signature, validity and revocation status as any relying party does, and records nothing but the pseudonym and the number of refills served to it in the epoch. It then signs the $k$ blinded tokens that the agent sends, following the publicly verifiable issuance protocol of Privacy Pass, and the agent unblinds and stores them~\cite{rfc9578,rfc9474}. The agent blinds against the issuer's key as published in the issuer list, so that no issuer can mark a person by offering them a key of their own. Because the signature is blind, the issuer learns nothing about the tokens it signs and cannot recognise one when it is later spent, even with the help of the site that received it.

\begin{figure}[ht]
  \centering
  \includegraphics[width=.75\textwidth]{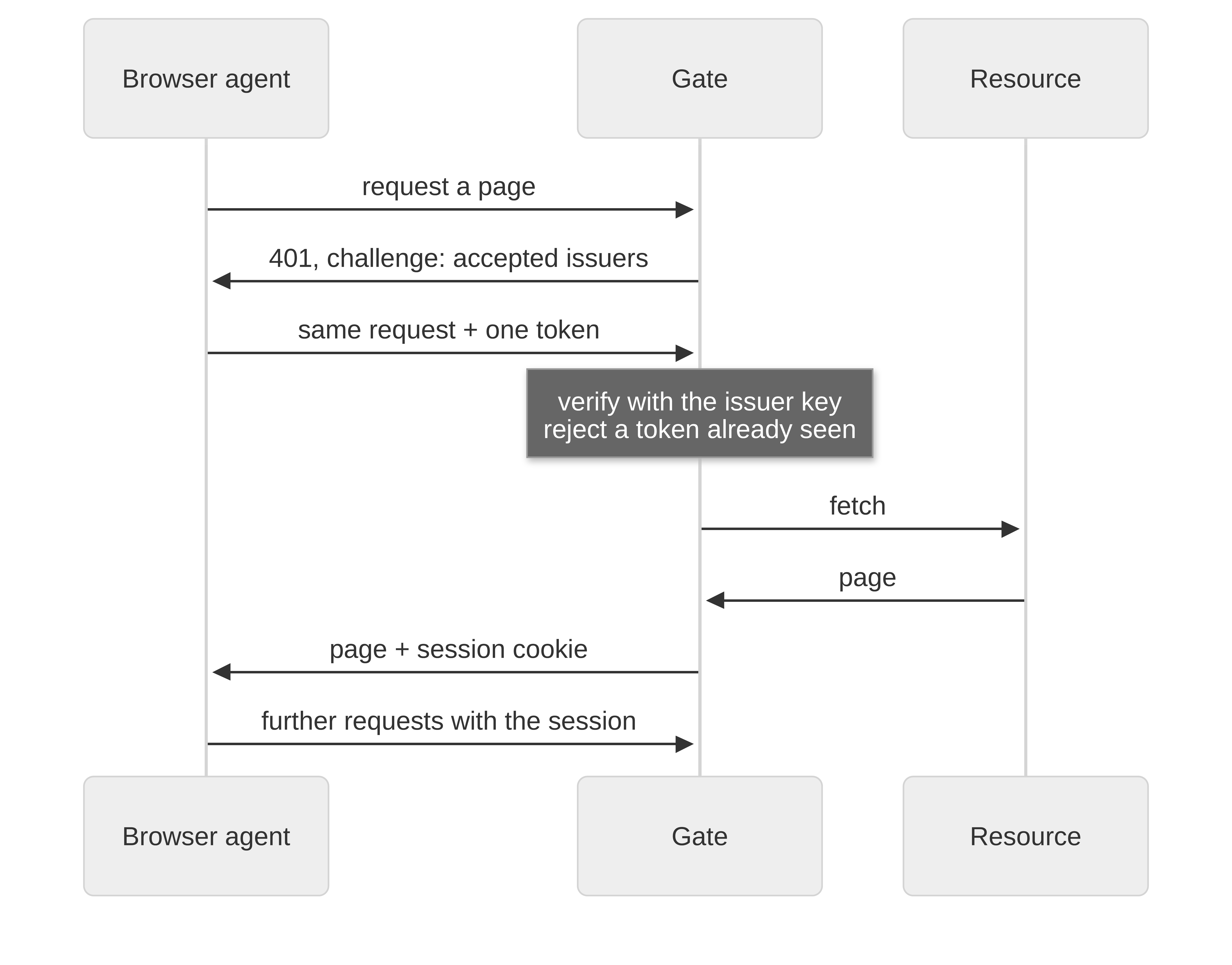}
  \caption{Visit: one token per site session.}
  \label{fig:visit}
\end{figure}

A visit, shown in Figure~\ref{fig:visit}, involves neither the wallet nor the issuer. The gate answers a request that carries no session with the standard Privacy Pass challenge: a response with status 401~\cite[\S15.5.2]{rfc9110} whose authentication header names the issuers the site accepts~\cite{rfc9577}. For each issuer it offers two challenges~\cite[\S2.1.1]{rfc9577}. One is bound to the site, for an agent that obtains a token on the spot. The other is the issuer's standing challenge, against which the agent minted the tokens it holds. Stored tokens are therefore not bound to a site: a token is valid at any gate that accepts its issuer, once per gate. The agent picks a token from an accepted issuer and repeats the request with the token attached. The gate verifies the token offline with the issuer's public key taken from the issuer list, and rejects a token it has already seen. It then opens a session for the origin with a cookie of bounded lifetime. Within the session the agent reads freely, at no further token, subject to a rate limit that the gate enforces with the status code for too many requests~\cite{rfc6585}.

The token is the publicly verifiable type of Privacy Pass, a blind RSA signature over a random value, and carries no attribute~\cite{rfc9578}. The epoch is the lifetime of an issuer key. The issuer rotates its key at the end of each epoch and publishes the new one in the issuer list. A token therefore expires with the key that signed it, and no revocation list is needed~\cite{rfc9576}. Each pseudonym obtains at most $k$ tokens per epoch from an issuer, all at the refill. The bound is a policy of the issuer, published in the issuer list, and is enforced against the pseudonym, never against a name. It bounds a person only if the wallet gives a person one pseudonym at the issuer, which the wallets of today do not. Because the signature is blind, the issuer cannot count tokens per site: $k$ bounds the sessions one pseudonym can open on any one site in an epoch, and the same tokens serve every site. Today, therefore, $k$ is also the number of site sessions a person can open in an epoch with one pseudonym, and an issuer sets it well above what reading needs.

Declared machines take a separate lane. A crawler that signs its requests under Web Bot Auth identifies its operator through a key published in a directory~\cite{webbotauth_httpsig}. The gate admits such requests according to the site's policy, for instance for search engines and archives, so the site stays visible to them. This lane is identified rather than anonymous, and opening it is the site's choice.

Registration stops at the issuer. No site talks to the wallet, so no site needs an access certificate. The implementing regulation on registration defines a wallet-relying party as a relying party that intends to rely upon wallet units to provide its services~\cite{cir2025_848}. A gate relies on tokens alone. We read this to mean that a site running a gate is not a wallet-relying party and need not register, as the age-verification solution already assumes for sites that verify age attestations~\cite{av_blueprint_spec}. The PID is used as issued, with no new attribute. A static site joins by placing the gate in front of its files, as a proxy or as an edge function. The agent can be a browser extension today and a native browser component in time, without changing the protocol, which is the direction of PACT, anonymous credentials native to the browser~\cite{pact2026}.

Admission is symmetric. A site of the Clear Human Web is published by a person, and the gate proves it with a badge. The publisher's agent asks an issuer to blind-sign the site's domain name under the issuer's badge key of the epoch~\cite{rfc9474}. The issuer does not see the domain and counts badges against the publisher's pseudonym, at most $k_b$ per epoch. The gate serves the badge in a response header. The agent verifies it against the issuer list and shows that the site is vouched for by a person, as browsers show a secure connection today.

The badge proves that a person holding a PID vouched for this domain in this epoch. It does not prove who the person is, that the person wrote the content, or that the content is honest. A badge expires with the key of its epoch and is renewed like a token. A publisher who vouches for sites they do not control spends their own quota on it, as a renter does with tokens.

A deployment is defined by five parameters:
\begin{itemize}[leftmargin=1.6em,itemsep=1pt,topsep=2pt]
  \item the number of tokens ($k$) a pseudonym obtains per epoch from one issuer, published by the issuer;
  \item the epoch, the lifetime of an issuer key;
  \item the session, the lifetime of a gate cookie, bounded by the site;
  \item the rate limit, requests per session and unit of time, set by the site;
  \item the issuer list: who maintains it, and how issuers enter and leave it.
\end{itemize}

\section{Analysis}  \label{sec:analysis}

This section states which of the five properties hold against each of the seven adversaries, for the construction of today and for the optimum, and then states the bound on identities that the construction enforces. Two assumptions hold throughout: connections between agent, gate and issuer are encrypted, as web connections are today, and the agent spends each token once. Issuers verify the PID as the construction prescribes, except for the issuer that colludes with a site, which may not. The badge adds nothing to what any party learns. It is a blind signature over a public domain name, verified by the agent against the issuer list, and the issuer counts badges as it counts tokens, against the pseudonym.

\subsection{Adversaries and Properties}  \label{sec:adversaries}

Each adversary is met with what the protocol hands it and nothing more. This section states, adversary by adversary, what that is and which property survives it. Table~\ref{tab:adversaries} collects the outcome for the construction of today and for the optimum. A full mark means that the property holds by the protocol alone, a dotted mark that it holds with a reserve, an empty mark that the adversary defeats it. One reserve is common to every row: today \prop{4} holds per pseudonym, since the wallet does not limit the pseudonyms a person holds at an issuer, as the next subsection shows.

\begin{table}[h]
  \centering
  \small
  \renewcommand{\arraystretch}{1.15}
  \caption{Properties satisfied against each adversary, today and in the optimum.}
  \label{tab:adversaries}
  \begin{tabular}{@{}l ccccc c ccccc@{}}
    \toprule
                                    & \multicolumn{5}{c}{Today} &       & \multicolumn{5}{c}{Optimum}                                                                                 \\
    \cmidrule(lr){2-6} \cmidrule(l){8-12}
    Adversary                       & P1                        & P2    & P3                          & P4    & P5    &  & P1                            & P2   & P3   & P4   & P5    \\
    \midrule
    Honest-but-curious site       & \yes                      & \half & \half                       & \half & \yes  &  & \yes                          & \yes & \yes & \yes & \yes  \\
    Colluding sites               & \yes                      & \half & \half                       & \half & \yes  &  & \yes                          & \yes & \yes & \yes & \yes  \\
    Site colluding with an issuer & \no                       & \half & \half                       & \no   & \yes  &  & \multicolumn{5}{c}{no issuer}                              \\
    The state                     & \yes                      & \no   & \yes                        & \half & \yes  &  & \yes                          & \yes & \yes & \yes & \yes  \\
    Renter                        & \half                     & \yes  & \yes                        & \half & \half &  & \half                         & \yes & \yes & \yes & \half \\
    Delegated program             & \yes                      & \yes  & \yes                        & \half & \yes  &  & \yes                          & \yes & \yes & \yes & \yes  \\
    Network observer              & \yes                      & \no   & \no                         & \half & \yes  &  & \yes                          & \yes & \yes & \yes & \yes  \\
    \bottomrule
  \end{tabular}
  {\footnotesize
    \begin{tabular}{@{}ccc@{}}
      \yes{} holds      &
      \half{} holds with a reserve &
      \no{} does not hold
    \end{tabular}}
\end{table}

\paragraph{Honest-but-curious site.}
An honest-but-curious site runs the protocol as specified and keeps everything it receives. What it receives is a token: the identifier of an issuer key, a random value chosen by the agent and the issuer's blind signature over it. It learns which issuer vouches for the visitor, and no more (\prop{2}). Two tokens of the same person are independent random values, so nothing in the protocol links one session to another, on this site or on any other (\prop{3}). What links them is the network address and the session cookie, whose scope is the session by definition. The site admits at most $k$ tokens of each accepted issuer per pseudonym and epoch (\prop{4}), and a program spending the person's tokens is indistinguishable from the person (\prop{5}). \prop{1} rests on the issuer.

A site that departs from this model has one move the protocol does not stop. Stored tokens are minted against the issuer's standing challenge, so the token a visitor spends on site A is valid, unspent, at every other gate that accepts the same issuer. A can present it at B and open a session there. The replay counts against the visitor's bound on B and links the visitor to nothing, since the token is a random value. The agent avoids it by answering the site-bound challenge, which the gate also offers, at the price of one refill, with its wallet presentation, at every first visit to a site. That refill hands the issuer the time of the visit, which is why the construction makes stored tokens the default.

\paragraph{Colluding sites.}
Colluding sites pool what each of them receives. Tokens spent on different sites share nothing but the identifier of the issuer key; the sites cannot tell whether two visitors are one person. Their visitor counts add up to a number that counts a person once per site and that no party can correct into a count of persons. The address is the only link, as for any two sites today.

\paragraph{Site colluding with an issuer.}
Here a site pools what it receives with an issuer that it accepts. The issuer holds, for each pseudonym, the PID presentation, which is recognisable across refills~\cite[\S10.1]{sdjwt_ietf}, and the times and number of its refills; the site holds tokens, times and addresses. Blindness means that the issuer cannot recognise a token it signed~\cite{rfc9474}, so the pair cannot link a token to a pseudonym by the protocol (\prop{2}, \prop{3}). It can correlate times: the anonymity set of a token is the set of pseudonyms refilled under the same key in the epoch, and a refill followed closely by a visit narrows it. The agent refills ahead of need, so the two are adjacent only when it chooses the site-bound challenge. The issuer has one more move, issuing to each pseudonym under a different key so that the key identifies the person. The agent closes it by blinding only against the key published in the issuer list. What the issuer can do unilaterally is sign without verifying, or beyond $k$: every site that accepts it then admits machines, or more than $k$ identities, until its key leaves the list, and the tokens die with the key. The remedy is the governance of the list. The optimum has no issuer, and this adversary reduces to the honest-but-curious site.

\paragraph{The state.}
The state issues the PID and may also run an issuer, or obtain the records of one. As PID provider alone, it sees neither refills nor visits. If the issuer checks revocation against a status list that it downloads as a whole, the provider learns at most that an issuer fetched the list. As an issuer, or with the help of one, the state sees the PID presentation and can match it against its own records: it then knows which persons hold proofs, and how often they refill. \prop{2} fails for enrolment. It holds for reading: the signature is blind, so visits remain invisible to the state as to any issuer, and \prop{3} and \prop{4} hold. The state can also stop a person. Revoking the PID stops the next refill, and the tokens already held remain valid until the key of the epoch expires: the lever exists and is bounded by the epoch. In the optimum the presentation is a zero-knowledge proof, so the state learns nothing but that a valid PID exists, and \prop{2} holds; revocation remains its lever.

\paragraph{Renter.}
A renter is a person who sells the use of their own tokens to a client the construction would not otherwise admit. The buyer's client is vouched for by a person holding a PID, so \prop{1} and \prop{5} hold to the letter: the construction cannot tell a rented token from a delegated one. What bounds renting is \prop{4}, under the two conditions of the bound stated below. The renter has $k$ tokens per issuer and epoch, each valid once per gate. Everything they can sell is therefore at most $k$ identities per site per epoch, drawn from the very tokens they would otherwise spend themselves. The tokens sold are unlinkable to the renter, as to anyone, so the renter is exposed only through the issuer, which sees their refills only. In the optimum, with $k = 1$, what a renter sells on a site is their own single appearance there for the epoch, and renter and buyer cannot be two visitors.

\paragraph{Delegated program.}
A delegated program is software that a person lets act within their own quota: a crawler, a reader, an assistant. It spends the person's tokens and is, at every party, the person. It is bounded by the same $k$ per site and epoch and by the rate limit of each session. The costs fall on the person: a session that the gate ends for exceeding the rate limit costs a token, and a program that spends the quota leaves none. A program that takes tokens without consent is a renter without a sale, bounded in the same way. Nothing in the construction distinguishes a page fetched by a person from one fetched by their program.

\paragraph{Network observer.}
A network observer sees the traffic between agent, gate and issuer without being any of them: an access provider, a transit network. Of encrypted traffic it sees addresses, destinations, sizes and times, and no token. It learns which addresses talk to an issuer, and therefore which hold proofs, and which addresses visit which gate; it links sessions by address. \prop{2} and \prop{3} fail at the network layer, today, exactly as they do for the web without the Clear Human Web; the token adds nothing to what the observer sees. In the optimum the connection runs over Tor, the address says nothing, and the observer learns only that someone uses Tor.

\subsection{The Bound on Identities}  \label{sec:bound}

Let $s$ be a site, $I_s$ the set of issuers it accepts, $p$ a person and $e$ an epoch. Let $N(p,s,e)$ be the number of distinct visitors that $p$ can be on $s$ during $e$, where a visitor is a session opened at the gate. Then
\begin{equation}
  N(p,s,e) \;\le\; \sum_{i \in I_s} k_i ,
  \label{eq:bound}
\end{equation}
where $k_i$ is the number of tokens per pseudonym and epoch that issuer $i$ publishes in the issuer list. The bound holds on two conditions: every issuer in $I_s$ is honest, and $p$ enrols with each of them under one pseudonym. With one accepted issuer, or one quota $k$ common to all, $N \le k$. In the optimum, $N \le 1$.

Bound~\eqref{eq:bound} follows from the construction. A session costs one token that the gate has not seen before. The gate accepts tokens of the issuers in $I_s$ only, verified against the keys of the epoch. Issuer $i$ signs at most $k_i$ tokens per pseudonym per epoch, counted at the pseudonym the wallet generates for it. Every token dies with the key that signed it. The two conditions are where the bound can give. An issuer that does not count, or does not verify, breaks it for every site that accepts it, as the site colluding with an issuer shows.

The second condition is not enforced by the wallets of today. The pseudonym is generated by the wallet unit for the issuer, and the current concept requires a wallet unit to let its user hold several pseudonyms at one relying party and to delete them~\cite[PA\_04, PA\_07]{arf_topic_e_rr}. The refinement round notes that a limit cannot be enforced while a user can delete a pseudonym and get another one~\cite[PA\_31]{arf_topic_e_rr}, and specifies no binding between the pseudonym and the PID presented with it~\cite[\S4.1.1]{arf_topic_e_rr}. Today, therefore, $k$ bounds a pseudonym, and a person pays one PID presentation for every further $k$. The issuer could count on the PID credential instead, which it sees at each refill. That would rest on the very linkability the construction otherwise avoids, and fails once the wallet presents each copy of a batch-issued PID only once~\cite{arf_topic_b_batch}. What enforces the second condition is the rate-limited pseudonym, which the ARF keeps as a requirement and deems necessary in the long run~\cite[\S4.1.2]{arf_topic_e_rr}.

The bound counts sessions. Within a session the agent fetches pages at the rate $R$ the site allows, for the lifetime $T$ of the session cookie. A person, or their program, therefore fetches at most $N \cdot R \cdot T$ pages per site per epoch. $R$ and $T$ are the site's choices, and the construction only makes them enforceable per person rather than per address. The bound holds per site. An honest agent spends each token once, so its sessions across all sites in an epoch also number at most $\sum_i k_i$. An agent that presents the same token at several gates reaches the per-site bound on each of them, at the cost of linking those sessions. The bound on one site therefore holds whatever the agent does, and the global one holds for the agent that protects its own unlinkability.

Each identity costs a share of a person's quota. Appearing once more on a site costs one of the $k_i$ tokens that the person has for that issuer in the epoch. This holds whether the person spends the token, delegates it, rents it out or loses it to theft. This is the economics of the construction. There is no identity on any site of the Clear Human Web that is not one of the $\sum_i k_i$ of some pseudonym for that epoch. With $k_i = 1$ there is nothing to sell but oneself.

Two parties can react to abuse, and each sees only its own side. The issuer sees refills and never a visit. Its one sanction is to refuse further refills to a pseudonym, judged on its refill pattern alone, and the tokens that pseudonym already holds stay valid until the epoch ends. A site sees sessions and never a pseudonym. Its levers are the session lifetime and the rate limit, and it can neither name a visitor nor refuse one for the rest of the epoch. The gap between the two, a site that cannot report and an issuer that cannot see, is what the per-site pseudonym of the optimum closes.

\section{Discussion}  \label{sec:discussion}

The analysis states what the construction guarantees. This section states what the construction of today does not guarantee, how far it stands from the optimum and what each step toward it requires, and what remains open.

\subsection{Limitations}  \label{sec:limitations}

Three limits hold at every step, the optimum included. The construction excludes whoever holds no PID in a wallet: residents of third countries, and anyone without the personal device the wallet runs on. This is a consequence of the root of trust. The same root gives the state a lever that no platform has: revoking a PID stops the holder's refills. The lever is bounded by the epoch and survives the optimum.

A token says that a person vouches for the client, and the person may be reading, or may have handed the client to a program. Adler et al.\ note the same limit for personhood credentials in general: a holder might still hand over a verified account to an agentic AI~\cite{adler2024phc}. The Clear Human Web accepts this by design (\prop{5}) and bounds it by $k$: a program has the identities of the person behind it, and no more.

The bound on identities holds under its two conditions; in practice it also depends on what an identity costs and on who will sell one, and the record of proof-of-personhood designs is a warning. Ohlhaver et al.\ document how Idena, which verified humans with synchronous tests, fractured into hidden pools, most run by puppeteers who at best paid participants for access to their keys; by May 2022, 23 entities controlled at least about 40\% of the accounts~\cite{Ohlhaver2024PoP}. Mehta finds a marketplace where AI agents post bounties for human workers, a third of them through programmatic channels, for tasks that include credential fraud and authentication circumvention at a median of 25 dollars~\cite{mehta2026hiring}. The Clear Human Web only prices renting, at one of $k$ identities per site and epoch.

Two further limits belong to the construction of today. Without Tor the address links sessions and reveals who talks to an issuer, as Table~\ref{tab:adversaries} records for the network observer. Today's construction inherits this from the web; the optimum removes it with Tor.

The issuer is the one party with a view of persons, and that view has two edges. It sees every refill, with its time, so an issuer that is the state, or shares its records with it, knows who holds proofs. A refill made at the moment of a visit narrows the anonymity set of the token for an issuer that compares notes with a site. Stored tokens, obtained ahead of need, are the construction's answer to the second edge; the first waits for zero-knowledge presentations. A related leak is coarser and harder to remove: the token names its issuer, and where issuers are national the issuer names a Member State. Hiding the issuer within a set accepted by the verifier is an active line of work~\cite{katz2026issuerhiding}, and TS13 lists issuer hiding among the items out of its scope~\cite{eudi_ts13_zk}.

\subsection{Toward the Optimum}  \label{sec:ladder}

The optimum is three steps away from the construction of today, and none of them touches the web side: each changes what the wallet can prove and how the agent derives what it shows. Table~\ref{tab:ladder} names, for each step, the mechanism, the property it gains with respect to the previous one, and what must exist before it can be deployed.

\begin{table}[h]
  \centering
  \scriptsize
  \renewcommand{\arraystretch}{1.2}
  \renewcommand{\tabularxcolumn}[1]{m{#1}}
  \caption{Approximations of the optimum, from what is deployable today to what the standards still lack.}
  \label{tab:ladder}
  \begin{tabularx}{\linewidth}{@{}c>{\raggedright\arraybackslash}m{2.1cm}*{3}{>{\raggedright\arraybackslash}X}@{}}
    \toprule
       & Step                                                                                                                                                                                                                                          & Mechanism & Property gained & What must exist \\
    \midrule
    A0 & Tokens in the browser
       & PID presented plainly to one registered issuer; blind RSA tokens, $k$ per pseudonym and epoch; stored in the agent, one per site session
       & \prop{1}, \prop{2}, \prop{3} and \prop{5} with the reserves of Table~\ref{tab:adversaries}; $k$ per pseudonym
       & Wallets~\cite{eidas2}; registration of the issuer under the regulation~\cite{cir2025_848}                                                                                                                                                                                                     \\
    \midrule
    A1 & One pseudonym per person; once-only PID copies
       & The wallet limits a person to one pseudonym at the issuer~\cite{arf_topic_e}; batch-issued copies of the PID, each presented once~\cite{arf_topic_b_batch}
       & \prop{4} per person: the second condition of the bound enforced by the wallet. The issuer holds no identifier of the person but the pseudonym it needs for $k$; refills of one person are linked by the pseudonym alone
       & The rate-limited pseudonym, required but not supported by the current concept~\cite{arf_topic_e_rr}; batch issuance of the PID in wallets, which does not protect against a PID provider that subverts the batch~\cite{invisible_traces_2026}                                                 \\
    \midrule
    A2 & Zero-knowledge presentation
       & A reusable proof of PID possession, bound to the device, within validity and not revoked, disclosing nothing~\cite{eudi_ts13_zk,eudi_ts14_mms}
       & \prop{2} complete: the issuer, or the site, learns only that a valid PID exists. Without a pseudonym there is no $k$ and \prop{4} is lost, which A3 restores
       & A zero-attribute circuit; a wallet that enables zero-knowledge proofs in remote presentation~\cite{eudi_wallet_core_zkp,eudi_ref_wallet_android}                                                                                                                                              \\
    \midrule
    A3 & Per-site pseudonyms and Tor (the optimum)
       & A pseudonym derived in the proof from the PID, the site and the epoch; the connection over Tor~\cite{tor2004}
       & \prop{4} with $k = 1$ and no issuer; a site can refuse a pseudonym for the epoch without learning who it is~\cite{nymble2007}; \prop{2} and \prop{3} against the network
       & Pseudonyms in the zero-knowledge circuit; Tor in the agent                                                                                                                                                                                                                                    \\
    \bottomrule
  \end{tabularx}
\end{table}

What is missing is concentrated in the wallet. A1 waits for the rate-limited pseudonym, kept as a requirement and not currently supported~\cite{arf_topic_e,arf_topic_e_rr}, for which constructions exist in the literature~\cite{mayrhofer2025pseudonyms}, and for batch issuance of the PID, a discussion topic of the ARF~\cite{arf_topic_b_batch}. A2 waits for a zero-attribute circuit, which the published Longfellow circuit is not~\cite{frigo2026ecdsa}, and for a wallet application that enables the proofs its core library offers~\cite{eudi_wallet_core_zkp,eudi_ref_wallet_android}. A3 waits for pseudonyms in the zero-knowledge specifications, which TS13 lists among the items out of its scope~\cite{eudi_ts13_zk}; Tor exists~\cite{tor2004}.

Three designs close to the Clear Human Web sit on this path at different points.

PACT and MoLE build the browser side of A2 and A3. They provide anonymous credentials native to the browser, and an endorsement in which the client proves in zero knowledge that it holds a valid mdoc from an accepted set of issuers~\cite{pact2026,mole2026arch,mole2026protocols}. Their anchor is a scarce signal rather than an identity document, as Table~\ref{tab:comparison} records. Their nullifier is derived from a credential-bound secret and the epoch, with the binding to the verifier's challenge kept apart from it, so one credential yields one nullifier per epoch~\cite{mole2026protocols}. A nullifier that does not depend on the site either identifies the client across sites or limits it to one site per epoch. The optimum derives the pseudonym from the site as well, which is what makes $k = 1$ per site compatible with reading everywhere.

x401 asks the wallet at every site, on an HTTP route of its own, and names proving personhood among its uses~\cite{x401spec}. In the EU every site then relies on wallet units and must register~\cite{cir2025_848}, and a plain presentation links the visitor across sites~\cite[\S10.1]{sdjwt_ietf}. The steps above change what the wallet proves and leave the sites unregistered.

The age-verification solution has the shape of A0 for one attribute~\cite{av_blueprint_spec}. The difference is in the attestation. An age attestation carries its attribute and is issued in a batch that the provider has seen. The provider is therefore in the position of the issuer with the credential in hand, and the subversion of batch-issued credentials applies to it~\cite{invisible_traces_2026}. A token carries nothing and was never seen by the issuer. This is why the Clear Human Web can afford a single issuer that knows the pseudonym: what it signs tells it nothing about where the signature will be spent.

\subsection{Open Problems}  \label{sec:open}

The first three problems belong to the construction of today; the last to the zero-knowledge steps, from A2 on.

\emph{Governance of the issuer list.} The list is the trust anchor of the construction and the fifth parameter of a deployment. Who maintains it, how an issuer enters it, and how fast an issuer that signs without verifying leaves it, decide how much a site colluding with an issuer can do. The age-verification solution answers the same question with a trust list; the right unit for personhood, an EU-level list, national lists or a list per site, has yet to be chosen.

\emph{Several wallet units per person.} The bound on identities counts pseudonyms, and the pseudonym is generated by the wallet unit~\cite{cir2024_2979}. The ARF allows a device to host more than one wallet instance, each part of a wallet unit of its own~\cite[\S4.3.2]{eudi_arb}, and a PID is issued to a wallet unit~\cite{cir2024_2977}. A person with several wallet units therefore has several pseudonyms at the same issuer. One pseudonym per person across wallet units would need a binding between the pseudonym and the PID, which the current concept does not specify~\cite[\S4.1.1]{arf_topic_e_rr}.

\emph{Deterrents to renting.} Two constructions on paper would raise the price of a rented identity without naming anyone. Rate-limited tokens whose owner is revealed after more than $N$ uses in one context give a per-site bound that no issuer needs to count~\cite{everlasting2025art}. Non-transferable tokens built from betrayability let a buyer steal the seller's secret, so that a seller who helps a buyer redeem risks the secret itself~\cite{kretzler2025betrayability}. Neither is a standard or part of Privacy Pass, and neither has been fitted to the issuance protocol of the construction without a new primitive in the wallet.

\emph{A zero-attribute circuit.} A2 needs a proof of PID possession that discloses nothing. The published Longfellow circuit discloses at least one attribute, and a zero-attribute variant is not described~\cite{frigo2026ecdsa}. Until one is, a disclosed attribute of no consequence may have to stand in for it.

\section{Conclusion}  \label{sec:conclusion}

One bit of information, that a person stands behind a client, is enough to let a part of the public web emerge wherever sites choose to join it. The EU wallet supplies that bit without a new root of trust: the PID each Member State issues once per person, shown to one issuer and carried to any site as a blind token. We have defined the Clear Human Web through five properties and seven adversaries, and shown which of them the construction of today satisfies.

With the Clear Human Web, whoever publishes has a fence that holds by cryptography where the Robots Exclusion Protocol only asks. Whoever measures has a count: a person is at most $k$ visitors on one site in an epoch, once the wallet limits pseudonyms, and one in the optimum. Nobody can add the visitors of two sites into a count of persons, since the same proof forbids linking. Whoever reads has proof: the publisher of each site holds the same proof as its readers.

Every step toward the optimum falls on the wallet, so a site that runs the gate need not change as the wallet improves. Two extensions follow from the same proof. Writing, as in comments, edits or votes, needs a visitor that a site can recognise and refuse without naming them, which is the per-site pseudonym of the optimum~\cite{nymble2007}. Search inside the Clear Human Web needs an engine that indexes only sites with a badge and serves only persons.

\renewcommand{\bibfont}{\small}
\setlength{\bibsep}{2pt}
\bibliographystyle{unsrtnat}
\bibliography{references}

\end{document}